# Secret Key Cryptosystem based on Non-Systematic Polar Codes

Reza Hooshmand, Mohammad Reza Aref, and Taraneh Eghlidos

*Abstract*—Polar codes are a new class of error correcting linear block codes, whose generator matrix is specified by the knowledge of transmission channel parameters, code length and code dimension. Moreover, regarding computational security, it is assumed that an attacker with a restricted processing power has unlimited access to the transmission media. Therefore, the attacker can construct the generator matrix of polar codes, especially in the case of Binary Erasure Channels, on which this matrix can be easily constructed.

In this paper, we introduce a novel method to keep the generator matrix of polar codes in secret in a way that the attacker cannot access the required information to decode the intended polar code. With the help of this method, a secret key cryptosystem is proposed based on non-systematic polar codes. In fact, the main objective of this study is to achieve an acceptable level of security and reliability through taking advantage of the special properties of polar codes. The analyses revealed that our scheme resists the typical attacks on the secret key cryptosystems based on linear block codes. In addition, by employing some efficient methods, the key length of the proposed scheme is decreased compared to that of the previous cryptosystems. Moreover, this scheme enjoys other advantages including high code rate, and proper error performance as well.

## I. Introduction

NOWADAYS, development and rapid dissemination of wireless communication systems have increased the demand for providing reliable and secure data. In this respect, channel coding is the study of techniques used for establishing a reliable communication between a sender and a receiver in the presence of channel errors. Cryptography is also known as the study of various methods employed to build secure communications in the presence of adversaries. In general, channel coding can be applied to provide two major categories of security; namely the information theoretic security and the computational security. Utilizing practical channel codes such as Low Density Parity Check (LDPC) codes [1] and Polar codes [2] in the structure of wiretap channel to achieve secrecy capacity is an instance of applying channel codes in establishing information theoretic security [3, 4]. In the same vein, taking advantages of various channel codes in the structure of public/secret key code based cryptosystems can be regarded as an application of channel coding in providing computational security [5, 6].

It is noteworthy that code based cryptosystems provide security and reliability in one process to guarantee the confidentiality and the integrity of the transmitted data. Besides, a combination of security and reliability in the structure of these systems can result in reducing the processing cost or providing a rather higher efficiency. Moreover, code based cryptosystems are considered as one of the important classes of cryptographic systems which are believed to resist quantum computers [7]. Establishing a suitable tradeoff between security and reliability is thus one of the important goals in designing such cryptosystems, which can be properly achieved through efficient linear codes employed in the structure of these cryptosystems.

The security of some code based cryptosystems is dependent upon the difficulty of the general decoding problem [8]. For an arbitrary binary linear code $C$, with a length of $N$ and dimension of $K$, for instance, the general decoding problem is that of decoding a channel output vector $\boldsymbol{y} = y_1^N = (y_1, y_2, \cdots, y_N)$ into the closest codeword $\boldsymbol{x} = x_1^N = (x_1, x_2, \cdots, x_N)$. In this case, the Hamming distance between $\boldsymbol{x}$ and $\boldsymbol{y}$, $d_H(\boldsymbol{x}, \boldsymbol{y}) = |\{i | 1 \leq i \leq N, x_i \neq y_i\}|$, is minimal [9]. It was earlier proved that the decoding problem of arbitrary linear codes belongs to the class of NP-complete problems [8].

*A. Related Works*

In 1978, McEliece proposed the first public key cryptosystem which was based on Goppa codes [5]. Compared with other public key cryptosystems, McEliece's cryptosystem enjoyed high speed encryption/decryption algorithms. However, this scheme had its own weaknesses such as low information rate and large key size. Later in 1984, the first secret key code based cryptosystem was suggested by Rao [10]. Although very similar to McEliece's cryptosystem, this scheme kept the public key secret. It was shown later that Rao's scheme could be broken by chosen plaintext attacks [6].

This work was supported in part by Iranian National Science Foundation (INSF) Cryptography Chair.

Reza Hooshmand is with the Department of Electrical Engineering, Science and Research Branch, Islamic Azad University, Tehran, Iran (e-mail: r.hooshmand@srbiau.ac.ir).

Mohammad Reza Aref is with the Department of Electrical Engineering, Sharif University of Technology, Tehran, Iran. (e-mail: aref@sharif.edu).

Taraneh Eghlidos is with the Electronics Research Institute, Sharif University of Technology, Tehran, Iran. (e-mail: teghlidos@sharif.edu).

In 1986, Rao and Nam introduced a modified secret key cryptosystem which allowed the use of short length Hamming codes with high information rate while improving the security level [6]. The modified scheme was called Rao-Nam (RN) cryptosystem. Not unlike the McEliece's cryptosystem, the security of RN scheme relies on the difficulty level at which the general linear codes can be decoded. Many modifications to RN scheme have already been proposed which are based on either applying various channel codes in its structure or modifying the set of allowed error vectors [11-15].

In the recent years, some efficient and secure secret key cryptosystems based on Turbo codes [1] and LDPC codes have been introduced. Turbo codes have also been employed in different secure channel coding schemes to be used in satellite communications [16, 17]. The issue of using quasi-cyclic low-density parity-check (QC-LDPC) codes in secret key cryptosystems is also addressed in [18, 19]. Due to the cyclic and sparse structure of the parity check matrix of QC-LDPC codes, the key lengths of these schemes were decreased significantly compared with previous RN-like schemes.

The idea of applying polar codes to provide information theoretic security has extensively been addressed in several researches [4, 20]. However, in spite of the interesting properties of the polar codes, these efficient codes have not been applied in the structure of cryptosystems based on general decoding problem. Recently, we introduced, for the first time to the best of our knowledge, the application of polar codes in the structure of secret key cryptosystem over binary erasure channel [21]. In fact, the present paper is a continuation and extension of our previous work in the context of secret key cryptosystems based on channel coding.

*B. Contributions of the proposed scheme*

The present paper is aimed at introducing a secret key cryptosystem which makes use of non-systematic finite length polar codes in an efficient way to overcome the problems arisen from insecure and unreliable communication channels. The proposed scheme is designed in such a way so as to avoid the weaknesses of the RN cryptosystem and is expected to provide more security and reliability. The main contribution of this work is the technique proposed for hiding the generator matrix of polar codes from the attacker. In fact, with the help of this method, the underlying cryptosystem can achieve a proper security level based on general decoding problem.

It has to be noted that the proposed scheme resists against the typical attacks on the cryptosystems based on channel coding. In addition, its error performance, key length and computational complexity will also be investigated to assess the efficiency. In order to evaluate the reliability of this scheme, the upper bound on error probability of the polar code used under Successive Cancelation (SC) decoding is being discussed in details as well. To decrease the key size of this scheme, we apply efficient techniques including, (1) utilizing the special structure of the generator matrix of polar codes, (2) using the efficient method based on pseudorandom number generator [22] to generate the nonsingular and permutation matrices, and (3) exploiting the non-systematic property of polar codes to generate the intentional error vectors. In fact, it is shown that the proper tradeoff between the security and reliability is attainable through the proposed scheme.

*C. Outline*

The rest of this paper is organized as the follows. Sections II & III give brief reviews of the polar codes and Rao-Nam cryptosystem, respectively. The concept of using polar codes in the structure of secret key cryptosystem is introduced in Section IV. The efficiency and security levels of the proposed cryptosystem are also assessed in Sections V & VI respectively. Finally, Section VII concludes the paper with a brief discussion of the future work.

## II. POLAR CODES

In this section, a brief description of the structure of polar codes will be presented and subsequently, an existing technique for constructing their generator matrix will be reviewed. Polar codes are a class of linear block codes that provably achieve the capacity of any symmetric Binary-input Discrete Memoryless Channel (B-DMC), such as BEC and Binary Symmetric Channel (BSC). Let $W : \mathcal{X} \to \mathcal{Y}$ be a B-DMC with input alphabet of $\mathcal{X} = \{0,1\}$, output alphabet of $\mathcal{Y}$ and transition probabilities of $\{W(y|x), x \in \mathcal{X}, y \in \mathcal{Y}\}$. Let us consider the following parameters for a B-DMC $W$ [2].

$$I(W) \triangleq \sum_{y \in \mathcal{Y}} \sum_{x \in \mathcal{X}} \frac{1}{2} W(y|x) \log \frac{W(y|x)}{\frac{1}{2}W(y|0) + \frac{1}{2}W(y|1)},$$

$$Z(W) \triangleq \sum_{y \in \mathcal{Y}} \sqrt{W(y|0)W(y|1)},$$

where $I(W) \in [0,1]$ is the mutual information between the input and the output of $W$ with uniform distribution on the input. When $W$ is a symmetric channel, $I(W)$ is called the capacity of $W$ and thus applied as the measure of rate. Besides, $Z(W) \in [0,1]$ is known as the Bhattacharyya parameter of $W$ and used as a criterion of measuring reliability. Note that $I(W) \approx 1$ iff $Z(W) \approx 0$, also $I(W) \approx 0$ iff $Z(W) \approx 1$. If $W$ is a BEC with erasure probability $\epsilon$, denoted by BEC($\epsilon$), then $Z(W) = \epsilon$ and $I(W) = 1 - Z(W) = 1 - \epsilon$ [2].

Let $\{W_N^{(i)} : 1 \leq i \leq N\}$ be a set of polarized binary input channels with indices '$i$' that can be obtained by performing a phenomenon on the $N$ independent copies of given B-DMC $W$. This phenomenon is called channel polarization and the polarized binary input channels are called bit-channels or sub-channels. By exploiting the channel polarization, the symmetric capacity terms $\{I(W_N^{(i)}), 1 \leq i \leq N\}$ and Bhattacharya parameters $\{Z(W_N^{(i)}), 1 \leq i \leq N\}$ of all $N$ bit-channels tend to 0 or 1 if $N$ is large enough [2]. In the remainder of this paper, the Bhattacharya parameter of $i$-th bit-channel, $Z(W_N^{(i)})$, is denoted by $Z_{N,i}$. Besides, we consider the methods which are proposed to obtain the Bhattacharya parameters of the bit-channels. Such parameters are necessary to construct the generator matrix of polar codes.

Let $\mathcal{J} = \{i, 1 \leq i \leq N\}$ be a set of all bit-channel indices. Let $A$ be a $K$-element subset of $\mathcal{J}$ which is called information set. Let $A^c$ be an $(N-K)$-element subset of $\mathcal{J}$ which is a complement to the subset $A$ and is called frozen (fixed) set. These sets are specified in such a way that $Z_{N,i} \leq Z_{N,j}$ for all $i \in A, j \in A^c$. In other words, it is possible to construct $N$ bit-channels such that their $NI(W)$ with indices in the information set tend to become reliable or noiseless and their $N(1 - I(W))$ with indices in the frozen set tend to become unreliable or noisy [2, 23].

### A. Constructing the Generator Matrix

Consider $N = 2^n, n \geq 1$ and $F = \begin{bmatrix} 1 & 0 \\ 1 & 1 \end{bmatrix}$. Given the rate $R < I(W)$ and the dimension $K = 2^n R$, a $K \times N$ generator matrix $G_A$ is constructed for any $(N, K)$ polar code through the following steps [24]:

1) Compute the $n$-th kronecker product $G_N = F^{\otimes n}$ which gives an $N \times N$ matrix. Then, label the rows of $G_N$ from top to bottom as $i = 1, 2, \cdots, N$.
2) Obtain the Bhattacharyya parameters of all $N$ bit-channels in the form of $Z_N = (Z_{N,i}, 1 \leq i \leq N)$ through the following recursive formula with initial condition $Z_{1,1}$.

$$Z_{2k,i} = \begin{cases} 2Z_{k,i} - Z_{k,i}^2 & 1 \leq i \leq k \\ Z_{k,i-k}^2 & k+1 \leq i \leq 2k \end{cases}, k = 1, 2, 2^2, \cdots, 2^{n-1}$$
(1)

If the channel $W$ is a BEC($\epsilon$), the initial condition $Z_{1,1}$ is equal to $\epsilon$.

3) Form a permutation $\pi_N = (i_1, \ldots, i_N)$ for the set of $N$ bit-channel indices $\mathcal{J} = \{1, 2, \cdots, N\}$ in such a way that the inequality $Z_{N,i_j} \leq Z_{N,i_k}, 1 \leq j < k \leq N$ is satisfied.
4) Obtain the information set $A \subset \mathcal{J}$ whose bit-channel indices correspond to $K$ leftmost indices of the permutation $\pi_N$, i.e. $i_1, \ldots, i_K$. Then, obtain the frozen set $A^c \subset \mathcal{J}$ whose bit-channel indices correspond to $N - K$ rightmost indices of the permutation $\pi_N$, i.e. $i_{K+1}, i_{K+2}, \ldots, i_N$.
5) Construct the generator matrix $G_A$ by choosing $K$ rows of the matrix $G_N$ which correspond to the bit-channel indices of the information set $A$. If the bit-channel $W_N^{(i)}$ is chosen, then the $i$-th row of $G_N$ is selected. Also, construct $(N-K) \times N$ matrix $G_{A^c}$ by selecting $N - K$ rows of $G_N$ corresponding to the bit-channel indices of the frozen set $A^c$.

In short, the Bhattacharya parameters $\{Z_{N,i}, 1 \leq i \leq N\}$ of all bit-channels $\{W_N^{(i)}, 1 \leq i \leq N\}$ are generated by recursive formula (1). Then, the generator matrix $G_A$ is constructed by choosing the $K$ rows of the matrix $G_N$ whose indices correspond to bit-channels with the least possible Bhattacharya parameters.

### B. Non-Systematic Encoding

Polar codes, introduced in [2], are in fact non-systematic codes. In systematic encoding, the information bits appear transparently as part of the codeword, while this is not the case in the non-systematic encoding. In the case of non-systematic polar codes with block length of $N$, an input vector $\boldsymbol{u} = (u_1, u_2, \cdots, u_N) = (\boldsymbol{u}_A, \boldsymbol{u}_{A^c})$ consists of two subvectors, namely the information vector, which is a $K$-bit subvector $\boldsymbol{u}_A = (u_i, i \in A)$ and the frozen (fixed) vector which is an $(N-K)$-bit subvector $\boldsymbol{u}_{A^c} = (u_i, i \in A^c)$. The information vector $\boldsymbol{u}_A$ comprises of information data that is free to change in each process of transmission, while the frozen vector consists of fixed values known to decoder [25]. In addition, the input vector $\boldsymbol{u}$ is encoded to $N$-bit codeword $\boldsymbol{x}$ as follows,

$$\boldsymbol{x} = \boldsymbol{u}_A G_A + \boldsymbol{u}_{A^c} G_{A^c} = \boldsymbol{u}_A G_A + c.$$

Since $c \triangleq \boldsymbol{u}_{A^c} G_{A^c}$ is a fixed vector, the encoder mapping $\boldsymbol{u}_A$ to $\boldsymbol{x}$ is non-systematic [25]. The code rate is defined as $R = |\boldsymbol{u}_A|/|\boldsymbol{x}| = |A|/N$ which can be adjusted by selecting the size of information set $A$. The coordinates of the information vector can be transmitted at a rate close to 1 through noiseless bit-channels. However, the coordinates of the frozen (fixed) vector can be transmitted at a rate close to 0 across the noisy bit-channels. Therefore, polar codes are efficient for channel coding [2].

### C. Successive Cancellation Decoding

Let $\boldsymbol{x}$ be an $N$-bit codeword of the polar codes which is transmitted across the $N$ bit-channels. Let $\boldsymbol{y}$ be the corresponding channel output vector which is decoded by the low complexity SC decoding algorithm. The main goal of the SC decoder is to obtain the estimated input vector by the knowledge of information set $A$, frozen vector $\boldsymbol{u}_{A^c}$ and as well as the channel output vector $\boldsymbol{y}$. The bits of input vector are estimated successively at the SC decoder in the following way [2],

$$\hat{u}_i = \begin{cases} u_i, & \text{if } i \in A^c \\ h_i(y_1^N, \hat{u}_1^{i-1}) & \text{if } i \in A \end{cases},$$

where decision functions $h_i: \mathcal{Y}^N \times \mathcal{X}^{i-1} \to \mathcal{X}, i \in A$, are computed as below for all $y_1^N \in \mathcal{Y}^N, \hat{u}_1^{i-1} \in \mathcal{X}^{i-1}$,

$$h_i(y_1^N, \hat{u}_1^{i-1}) \triangleq \begin{cases} 0, & \text{if } \frac{w_N^{(i)}(y_1^N, \hat{u}_1^{i-1}|0)}{w_N^{(i)}(y_1^N, \hat{u}_1^{i-1}|1)} \geq 1 \\ 1, & \text{otherwise} \end{cases}.$$

The information bits $u_i, i \in A$, are estimated one by one using the $i$-th decision element after the channel output vector $\boldsymbol{y}$ and the previous estimated information bits $\hat{u}_1^{i-1}$ are known. Furthermore, the value of frozen bits, $u_i, i \in A^c$, is known to the SC decoder. It has been proved that for any given B-DMC $W$, the error probability under SC decoding is upper bounded as follows [2],

$$P_e \leq \sum_{i \in A} Z_{N,i}. \tag{2}$$

Also, it has been indicated that reliable communication using SC decoder is obtained when the following relation is satisfied [26, 27],

$$R < I(W) - N^{-1/\mu}, \quad (3)$$

where $\mu$ is the scaling exponent, whose values depends on the choice of channel. Its value for BEC, for instance, is approximately equal to 3.627. Indeed, reliable communication under SC decoding for any B-DMC $W$ is obtained when the rate is less than the capacity at least to the extent of $N^{-1/\mu}$. It can be considered as a tradeoff between the rate and the block length of polar codes for a given error probability, when the SC decoder is utilized [27]. In this paper, the maximum possible code rate fulfilling (3) is named by *cutoff rate* and denoted by $R_0$.

### III. THE RAO-NAM CRYPTOSYSTEM

The Rao-Nam (RN) cryptosystem is an important secret key code based cryptosystem used as a reference to measure the security and efficiency of secret key cryptosystems based on error correcting codes. In this section, the structure of RN scheme is being described, followed by an in-depth investigation of its drawbacks.

*A. Secret Key*

The secret key of the RN scheme consists of the parameters $\{G, S, P, \mathcal{T}\}$ which are explicated as follows [6]:
1) Let $G$ be a $K \times N$ generator matrix of the binary linear code $C$.
2) Let $S$ be a $K \times K$ random binary nonsingular matrix (scrambler).
3) Let $P$ be an $N \times N$ random binary permutation matrix (permutor).

In RN cryptosystem, a set of predetermined $N$-bit intentional error vectors, $\mathcal{E} = \{e_i, 1 \leq i \leq \mathcal{N}_e\}$, with cardinality $\mathcal{N}_e = 2^{N-K}$ is considered which has two main properties. The first property, called the weight property, is that all error vectors have the average Hamming weight equal to half of the code length, $w_H(e) \approx N/2$. The second property, i.e. the syndrome property, is that no distinct error vector is located in the same coset of $C$ [28]. According to these definitions, the syndrome error table can be defined as follows.

4) Let $\mathcal{T} = \{e_i H^T | e_i \in \mathcal{E}\}$ be a predetermined set of error vectors which is also called the syndrome error table. This set consists of $2^{N-K}$ cosets each of which has a distinct syndrome $s_i = e_i H^T$. Therefore, any set of $N$-bit error vectors can be selected, one from each of $2^{N-K}$ cosets.

*B. Encryption*

A $K$-bit message $m = (m_1, m_2, \cdots, m_K)$ is encrypted into an $N$-bit ciphertext $c = (c_1, c_2, \cdots, c_N)$ as shown below [6].

$$c = (mSG + e)P = mG' + eP, \quad G' = SGP,$$

where $G'$ is a $K \times N$ encryption matrix. Besides, $e$ denotes an $N$-bit intentional error vector selected randomly from the syndrome error table $\mathcal{T}$.

*C. Decryption*

A ciphertext $c$ is decrypted into a plaintext $m$ using secret keys $S^{-1}$, $H^T$ and $P^T$ following the steps below [6].
1) Compute $c' = cP^T = mSG + e = m'G + e$, $m' = mS$.
2) Calculate the syndrome $s = c'H^T = m'GH^T + eH^T = eH^T$, $GH^T = 0$. Find the corresponding error vector $e$ from the syndrome error table $\mathcal{T}$.
3) Obtain $m'G = c' - e$ and recover $m'$ using the decoding algorithm.
4) Multiply $m'$ by $S^{-1}$ to retrieve the message $m$.

*D. Weaknesses*

The RN scheme has several drawbacks as being discussed below:
1) One of disadvantages of the RN scheme is that it needs to store the matrices $S, P$ and $G$. Similarly, the syndrome error table $\mathcal{T}$ should be saved to remove the errors in the decryption process. Therefore, a large amount of secret keys are exchanged and stored by both the sender and the receiver [6].
2) Yet another practical problem of this scheme lies in the small number of error vectors for their recommended code parameters, e.g. $\mathcal{N}_e = 2^{N-K} = 2^8$ for $(72, 64)$ Hamming code. Hence, the RN scheme is vulnerable to chosen plaintext attacks [6]. Another drawback is the possibility of estimating the rows of encryption matrix $G'$ of this scheme using the majority voting analysis [28, 29].
3) In RN scheme, there exists a tradeoff between the code rate and the security. In fact, the code length $N$ is impractical for having a high code rate and a large number of intentional error vectors [11]. Furthermore, the RN scheme preserves the error correction capability of the employed code only partially [30].

Given the mentioned shortcomings, this research attempts to address these problems through applying the interesting properties of non-systematic polar codes and other efficient methods.

### IV. THE PROPOSED CRYPTOSYSTEM

The proposed secret key cryptosystem is designed based on finite length polar codes so that channel errors are corrected and the information is concealed from an unauthorized user. To this end, we consider the transmission over a BEC($\epsilon$), as it has been shown that among all the B-DMCs $W$, the best tradeoff between rate and reliability belongs to BEC. In other words, for a BEC, the Bhattacharyya parameter $Z(W)$ is minimized among all channels with a given capacity $I(W) = 1 - Z(W)$. Besides, given the general B-DMCs, no efficient algorithm has been introduced so far to calculate the Bhattacharya parameters. For a BEC, however, these parameters are constructed efficiently using (1) [2].

Therefore, unlike the other B-DMCs, the method used for constructing polar codes is simple for the BECs and can thus be performed with a complexity of $\mathcal{O}(N)$ [24].

*A. A technique for hiding the generator matrix of polar codes*

In the computational security, it is assumed that the attacker has unlimited access to the transmission channel. Moreover, the generator matrix of the polar codes has a channel dependent structure. This can imply that the attacker can specify the generator matrix of these codes using the channel parameters, length and dimension of the intended polar code. The main question being addressed in the currecnt study is that of how to keep the generator matrix of polar codes secret from the attacker to use these efficient codes in the structure of cryptosystems based on general decoding problem. In response to this issue, an efficient method is being proposed here, through which, an attacker cannot construct the hidden generator matrix of polar codes over BEC($\epsilon$) even if the parameters $N, R$ and $\epsilon$ are known. Let's consider the set of $N$ bit-channel indices $\mathcal{I} = \{1, 2, \cdots, N\}$, the permutation $\pi_N = (i_1, i_2, \cdots, i_{NR_0}, \cdots, i_N)$ and the cutoff rate $R_0$ for an $(N, K)$ polar code, as defined in Section II.

*Remark 1.* The $NR_0$ bit-channels are regarded as *Good bit-channels* if the corresponding Battacharya parameters are minimized (i.e. the least error probability) among all $N$ bit-channels. That is, the indices of good bit-channels in the set $\mathcal{I}$ correspond to the indices $\{i_1, i_2, \cdots, i_{NR_0}\} \subset \pi_N$.

*Remark 2.* The $N(1 - R_0)$ bit-channels are regarded as *Bad bit-channels* if the corresponding Battacharya parameters are maximized (i.e. the most error probability) among all $N$ bit-channels. That is, the indices of bad bit-channels in the set $\mathcal{I}$ correspond to the indices $\{i_{NR_0+1}, i_{NR_0+2}, \cdots, i_N\} \subset \pi_N$.

The following section explains how the generator matrix of polar codes can be kept secret.

1) Consider the method of constructing the generator matrix for an $(N, K)$ polar code as discussed in Section II-A. First, all Bhattacharya parameters of $N$ bit-channels, $Z_{N,i}, 1 \leq i \leq N$, and the permutation $\pi_N$ are constructed. Now, in order to keep the generator matrix secret, $K$ indices are selected randomly from the indices of good bit-channels. Indeed, this step is equivalent to the random selection of $K$ bit-channels from $NR_0$ good bit-channels. Subsequently, the randomly selected indices of the set $\mathcal{I}$ are considered as the *secret information set*, denoted by $A(s)$. In fact, $A(s)$ is the subset of $\mathcal{I}$ with $K$ randomly selected indices of good bit-channels.

    The *secret generator matrix* $G_{A(s)}$ is defined as a $K \times N$ submatrix of $G_N$ whose $K$ rows are chosen in accordance with the indices of $A(s)$. If the cutoff rate $R_0$, the length $N$, and the dimension $K$, are selected properly, the number of polar codes equivalent to the used code is large enough. In this case, an attacker cannot obtain the secret generator matrix in polynomial time. However, as it is discussed in Section V-A, most probably, this selection is not the best choice to achieve channel capacity. Indeed, there is a tradeoff between the security and efficiency which is usually inevitable in cryptosystems based on channel coding.

2) The *secret frozen (fixed) set*, denoted by $A^c(s)$, is a subset of $\mathcal{I}$ whose elements are the $N - K$ non-selected indices of the set $\mathcal{I}$ in step 1. Moreover, the secret matrix $G_{A^c(s)}$ is defined as an $(N - K) \times N$ submatrix of $G_N$ whose $N - K$ rows are chosen based on the indices of the secret frozen set.

3) In order to have a more secure decoding process, the frozen vector should be concealed from an adversary. Since the polar code performance is not sensitive to the manner in which the frozen vector is selected, it makes no big difference how this vector is chosen [2]. Therefore, in the encryption/decryption process of the proposed scheme, an $(N - K)$-bit randomly chosen vector is generated by an $(N - K)$-bit LFSR to be used as the secret frozen vector, denoted by $\mathbf{u}_{A^c(s)}$. As a result, the number of possible frozen vectors is equal to $\mathcal{N}_F = 2^{N-K} - 1$. As long as the length and dimension of the employed polar code are selected properly, the attacker cannot find the secret frozen vector in polynomial time.

The inputs to SC decoder of polar codes are the channel output vector, the information set and the frozen vector. Hence, by hiding the information set and the frozen vector using the above technique, the attacker cannot decode the channel output vector $\mathbf{y}$ to the estimated input vector $\hat{\mathbf{u}}$ in polynomial time. Fig. 1(a, b) represents the proposed concept for providing security based on polar codes.

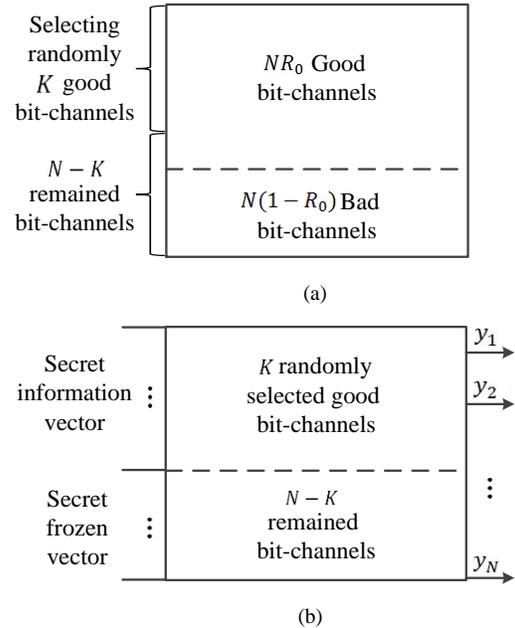

Fig. 1. The proposed concept for providing security based on polar codes, (a). The $K$ bit-channels are selected randomly from $NR_0$ good bit-channels, (b). The secret information vector is transmitted through $K$ randomly selected good bit-channels. Also, the secret frozen vector is transmitted across $N - K$ remained bit-channels

It is observable from Fig. 1(a) that, in order to hide the generator matrix $G_{A(s)}$, the $K$ bit-channels are selected randomly from $NR_0$ good bit-channels. In Fig. 1(b), the secret information vector, denoted by $\mathbf{u}_{A(s)}$, is transmitted through the $K$ randomly selected good bit-channels. Besides, it can be viewed that the secret frozen vector is transmitted across the $N - K$ remained bit-channels. In this case, should the parameters $N$, $K$ and $R_0$ are selected properly, an attacker cannot recognize on which bit-channels the secret information vector is transmitted. Thus the secret generator matrix $G_{A(s)}$ cannot be constructed by the attacker even if the transmission channel parameters, the length and dimension of the utilized polar code are known.

*B. Secret Key*

The set of keys which should be kept secret is $\mathcal{K} = \{G_{A(s)}, S, P, \mathcal{E}\}$. In this set, $G_{A(s)}$ is a $K \times N$ generator matrix of polar codes requiring $KN$ bits of memory, $\mathcal{E}$ is a set of $N$-bit intentional error vectors requiring $N|\mathcal{E}|$ bits of memory, $S$ is a $K \times K$ random binary nonsingular matrix and $P$ is an $N \times N$ random binary permutation matrix which require $K^2$ and $lg_2^{N!}$ bits of memory, respectively. By saving the set $\mathcal{K}$, the key length of the proposed scheme enlarges. Therefore, efficient methods are applied to reduce the size of the exchanged key dramatically. In this case, the secret key set is $\mathcal{K} = \{A^c(s), IV_S, IV_P, IV_F\}$ which consists of the parameters as follows:

1) As mentioned before, the secret generator matrix of an $(N,K)$ polar code is defined as the $K \times N$ submatrix of $G_N$ whose rows are chosen based on the indices of the secret information set $A(s)$. Hence, it will suffice to store only $A(s)$ instead of $G_{A(s)}$. On the other hand, since the secret frozen set is complement to secret information set and requires less memory to save, so it is possible to store $A^c(s)$ instead of $A(s)$.
2) Let $IV_S$ be a $(2K - 4)$-bit initial vector to generate a binary pseudorandom sequence $r_1, r_2, \cdots, r_{2K-4}, 0, 0$ by a $(2K - 2)$-bit LFSR. The generated pseudorandom sequence is used to construct the binary nonsingular matrix $S_{K \times K}$ (see Section V-B for more details).
3) Let $IV_P$ be an $(N - 2)$-bit initial vector of LFSR to generate a binary pseudorandom sequence $r_1, r_2, \cdots, r_{N-2}, 0$ by an $(N - 1)$-bit LFSR. The generated pseudorandom sequence can be used to construct the binary permutation matrix $P_{N \times N}$ (see Section V-B for more details).
4) Let $IV_F$ be an $(N - K)$-bit initial vector to generate an $(N - K)$-bit vector by an $(N - K)$-bit LFSR. Due to the non-systematic property of the employed polar code, the generated vector is used as secret frozen vector $\mathbf{u}_{A^c(s)}$. Thus, $\mathbf{e} = \mathbf{u}_{A^c(s)} G_{A^c(s)}$ can be considered as an $N$-bit intentional error vector and $\mathcal{E} = \{\mathbf{e}_i, 1 \leq i \leq \mathcal{N}_e\}$ with cardinality $\mathcal{N}_e = 2^{N-K}$ as a set of $N$-bit intentional error vectors. Apparently, unlike the RN cryptosystem, there is no need to store the syndrome error table $\mathcal{T}$.

It will be illustrated later in Section VI that reducing the key size of the proposed scheme by these efficient methods does not decrease the security level of the system.

*C. Encryption*

1) The sender first randomly chooses a code in a family of equivalent $(N,K)$ polar codes by selecting $K$ indices at random from $NR_0$ indices of good bit-channels. Then, the sender generates an $(N - K)$-bit frozen vector randomly using an LFSR with the initial value $IV_F$. In order to perform the decryption process properly, it is necessary to synchronize the sender and the receiver. This way, the frozen vector employed by the sender is known to the receiver synchronously. Subsequently, an intentional error vector $\mathbf{e}$ is constructed.
2) Finally, each $K$-bit message $\mathbf{m}$ is encrypted into an $N$-bit ciphertext $\mathbf{c}$ as shown below.

$$\begin{aligned}\mathbf{c} &= \left(\mathbf{m} S G_{A(s)} + \mathbf{u}_{A^c(s)} G_{A^c(s)}\right) P \\ &= \mathbf{m} S G_{A(s)} P + \mathbf{u}_{A^c(s)} G_{A^c(s)} P \\ &= \mathbf{m} G' + \mathbf{e} P,\end{aligned} \quad (4)$$

where $G' = S G_{A(s)} P$ is a $K \times N$ encryption matrix equivalent to the generator matrix $G_{A(s)}$.

*D. Decryption*

The ciphertext $\mathbf{c}$ is transmitted over the insecure channel and the channel output vector $\mathbf{y} = \mathbf{c} + \mathbf{e}_{ch} = \mathbf{m} G' + \mathbf{e} P + \mathbf{e}_{ch}$ is decrypted by the authorized receiver as described below.

1) The transposed permutation matrix, $P^T$, is multiplied by the channel output vector $\mathbf{y}$ and $\mathbf{y}' = {y'}_1^N = \mathbf{y} P^T = \mathbf{m} S G_{A(s)} + \mathbf{e} + \mathbf{e}_{ch} P^T$ is computed to remove the permutation matrix $P$. In this case, $\mathbf{e}_{ch} P^T$ is a vector having the same Hamming weight as $\mathbf{e}_{ch}$.
2) The authorized receiver makes use of the secret initial value $IV_F$ to generate the secret frozen vector. Then, the set $\{A(s), \mathbf{u}_{A^c(s)}, \mathbf{y}'\}$ is considered as the input to the SC decoder. Finally, the input vector $\mathbf{u} = (\mathbf{u}_{A(s)}, \mathbf{u}_{A^c(s)}) = (\mathbf{m} S, \mathbf{u}_{A^c(s)})$ is estimated by the SC decoder as:

$$\hat{u}_i = \begin{cases} u_i, & if\ i \in A^c(s) \\ h_i({y'}_1^N, \hat{u}_1^{i-1}) & if\ i \in A(s) \end{cases},$$

where the decision function $h_i: \mathcal{Y}^N \times \mathcal{X}^{i-1} \to \mathcal{X}$, $i \in A(s)$, is defined as:

$$\hat{u}_i = h_i({y'}_1^N, \hat{u}_1^{i-1}) \triangleq \begin{cases} 0, & if\ \frac{w_N^{(i)}({y'}_1^N, \hat{u}_1^{i-1}|0)}{w_N^{(i)}({y'}_1^N, \hat{u}_1^{i-1}|1)} \geq 1 \\ 1, & otherwise \end{cases}, i \in A(s).$$

3) Having obtained the secret information vector $\mathbf{u}_{A(s)} = \mathbf{m} S$ using the SC decoder, we can now recover the message as $\mathbf{m} = \mathbf{u}_{A(s)} S^{-1}$.

The secret information set $A(s)$ and secret frozen vector $\mathbf{u}_{A^c(s)}$ are necessary to initiate the SC decoder. Therefore, it is computationally infeasible for any unauthorized user to correct channel errors without the knowledge of parameters $(A(s), \mathbf{u}_{A^c(s)})$. Fig. 2 illustrates the block diagram of the proposed cryptosystem. As can be viewed from this figure, at the first step, the message is multiplied by the nonsingular matrix $S$. Then the $K$-bit secret information vector $\boldsymbol{u}_{A(s)} = \boldsymbol{m}S$ is encoded to the $N$-bit codeword $\boldsymbol{x} = \boldsymbol{u}_{A(s)}G_{A(s)} + \boldsymbol{u}_{A^c(s)}G_{A^c(s)}$. Eventually, the $N$-bit ciphertext $\boldsymbol{c} = \boldsymbol{x}P$ is obtained through multiplying the codeword $\boldsymbol{x}$ by the permutation matrix $P$.

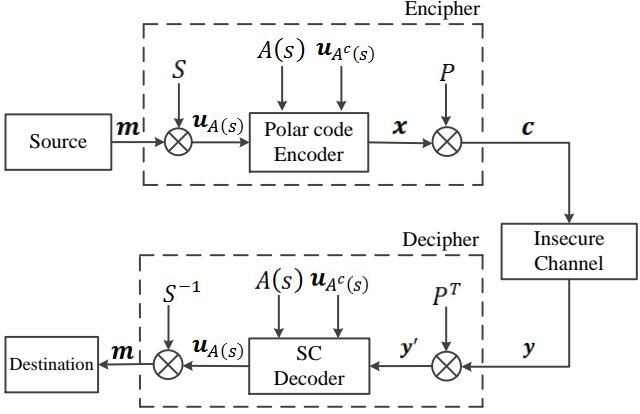

Fig. 2. Block diagram of the proposed cryptosystem.

The received vector $\boldsymbol{y}$ is also arrived at by transmitting the ciphertext through an insecure channel, which is then multiplied by the transposed permutation matrix. In the next step, the vector $\boldsymbol{u}_{A(s)}$ is obtained by performing the SC decoding on the $N$-bit vector $\boldsymbol{y}' = \boldsymbol{y}P^T$. Finally, the message $\boldsymbol{m}$ is recovered by multiplying the vector $\boldsymbol{u}_{A(s)}$ by the inverse of the nonsingular matrix $S$.

## V. Efficiency

The efficiency of the proposed cryptosystem is evaluated in terms of its error performance, key length and computational complexity. A detailed account of the observations is being provided below.

### A. Error Performance

The error performance of the used finite length polar codes is being analyzed under SC decoding. Yet the following remarks are to be taken into consideration first:

*Remark 3.* Let $A_1$ be a subset of $\mathcal{I} = \{1, 2, \cdots, N\}$ whose elements correspond to the indices $\{i_1, i_2, \cdots, i_K\} \subset \pi_N$. The *minimum upper bound on error probability* under SC decoding is equal to the sum of Battacharya parameters of $K$ bit-channels whose indices are the elements of the subset $A_1$, i.e. $P_{e_1} = \sum_{i \in A_1} Z_{N,i}$.

It has to be noted that this upper bound is the same as the standard upper bound on error probability of polar codes under SC decoding [2].

*Remark 4.* Let $A_2$ be a subset of $\mathcal{I} = \{1, 2, \cdots, N\}$ whose elements correspond to the indices $\{i_{NR_0-K+1}, i_{NR_0-K+2}, \cdots, i_{NR_0}\} \subset \pi_N$. The *maximum upper bound on error probability* under SC decoding is equal to the sum of the Battacharya parameters of $K$ bit-channels whose indices are the elements of the subset $A_2$, i.e. $P_{e_2} = \sum_{i \in A_2} Z_{N,i}$.

In the proposed scheme, since $K$ bit-channels are selected randomly from $NR_0$ good bit-channels, the upper bound on the error probability can vary from $P_{e_1}$ to $P_{e_2}$ depending on the sum of the Battacharya parameters of $K$ selected good bit-channels. In the sequel of this section, it will be discussed how some parameters such as erasure probability $\epsilon$, code length $N$, code rate $R$, and the manner in which the secret information set $A(s)$ is selected can affect $P_{e_1}$ and $P_{e_2}$.

If the transmission channel is BEC($\epsilon$), the initial value of the recursive formula (1) will be $Z_{1,1} = \epsilon$. Therefore, the erasure probability $\epsilon$ should be considered such that reliable communication is achieved. In this work, we consider the condition $P_{e_2} \leq 10^{-\alpha}$ to have a reliable communication where $\alpha$ has different values, depending on the application of the proposed scheme. Here, we select $\alpha = 4$ and based on which the analysis of the error performance is subsequently carried out. The erasure probabilities of BEC should be considered in a way that $P_{e_2}$ is less than or equal to $10^{-4}$. In this case, $P_{e_1}$ is definitely less than $10^{-4}$. As shown in table I, erasure probability varies in different intervals depending on the code lengths to satisfy the condition $P_{e_2} \leq 10^{-4}$.

TABLE I
DIFFERENT INTERVALS ON ERASURE PROBABILITY TO SATISFY $P_{e_2} \leq 10^{-4}$

| $N$ | $\epsilon$ |
|---|---|
| $2^9$ | $[0, 0.06]$ |
| $2^{10}$ | $[0, 0.07]$ |
| $2^{11}$ | $[0, 0.08]$ |
| $2^{15}$ | $[0, 0.12]$ |
| $2^{20}$ | $[0, 0.17]$ |

It is obvious that for larger code lengths, we can provide larger intervals on $\epsilon$ to achieve reliable communication. In addition, the code rate should be chosen in a way that $R < R_0$. In this scheme, in order to obtain a secure and reliable communication, finite length polar codes with high rate are employed. For instance, we consider a $(1024, 832)$ polar code with $R = 0.8125$ over BEC(0.01). Note that for BECs with larger or smaller $\epsilon$, it is possible to select other code rates depending on the application. For example, we will have $R > 0.83$ if $\epsilon < 0.01$ for the fixed block length $N = 2^{10}$.

Fig. 3 presents the variations of $P_{e_1}$ and $P_{e_2}$ in terms of $R \in [0.55, R_0]$. The polar code of length $N = 2^{10}$ is considered over BECs with $\epsilon = 0.01, 0.05, 0.1$. Two sets of three curves are depicted in this figure. The solid and the dashed lines plot $P_{e_1}$ and $P_{e_2}$ vs. code rate, respectively. As is evident in this figure, $P_{e_1}$ depends on the variations of the rate and erasure probability. Furthermore, the cutoff rate $R_0$ is increased as the erasure probability $\epsilon$ is decreased. It is also viewed that the cutoff rate $R_0$ is equal to 0.75, 0.8 and 0.84 for the erasure probabilities of 0.1, 0.05 and 0.01, respectively. This signifies the possibility to achieve reliable communication at higher code rates by increasing the cutoff rate $R_0$.

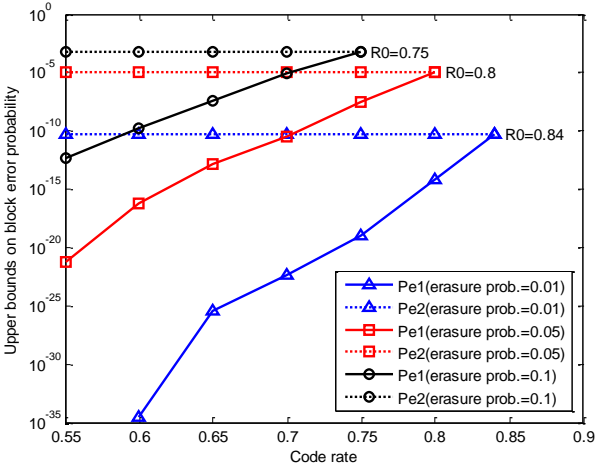

Fig. 3. Variations of $P_{e_1}$ and $P_{e_2}$ vs. rates $R \in [0.55, R_0]$ for the polar code of length $N = 2^{10}$ over BECs with $\epsilon = 0.01, 0.05, 0.1$.

On the other hand, $P_{e_2}$ is rate-independent. The main reason behind this is that the Battacharya parameters corresponding to set $A_1$ are rather small (approximately close to zero) compared to the ones corresponding to set $A_2$. Therefore, unlike the upper bound $P_{e_1}$, the value of $P_{e_2}$ is invariable in terms of rates. For the $(1024, 832)$ polar code over BEC(0.01), the upper bound on the error probability can vary from $P_{e_1} = \sum_{i \in A_1} Z_{1024,i} \approx 3.678 \times 10^{-13}$ to $P_{e_2} = \sum_{i \in A_2} Z_{1024,i} \approx 5.554 \times 10^{-11}$. According to remarks 3 and 4, $A_1$ and $A_2$ are the subsets of $\mathcal{I} = \{1, 2, \cdots, 1024\}$ whose elements correspond to the indices $\{i_1, i_2, \cdots, i_{832}\} \subset \pi_{1024}$ and $\{i_{29}, i_{30}, \cdots, i_{860}\} \subset \pi_{1024}$, respectively.

The code length $N$ is another parameter affecting $P_{e_1}$ and $P_{e_2}$ in the proposed scheme. Fig. 4 depicts the variations of $P_{e_1}$ and $P_{e_2}$ in terms of the rates $R \in [0.5, R_0]$ for the polar codes of lengths $N = 2^{10}, 2^{15}$ over BEC(0.05). It is observable that both $P_{e_1}$ and $P_{e_2}$ are decreased as the code length is increased. Further, the cutoff rate $R_0$ increases as the code length is enlarged. As can be seen, for the lengths $N = 2^{10}, 2^{15}$, the cutoff rate $R_0$ is equal to 0.8 and 0.89, respectively. In other words, it is possible to achieve reliable communication at higher code rates through increasing the code length.

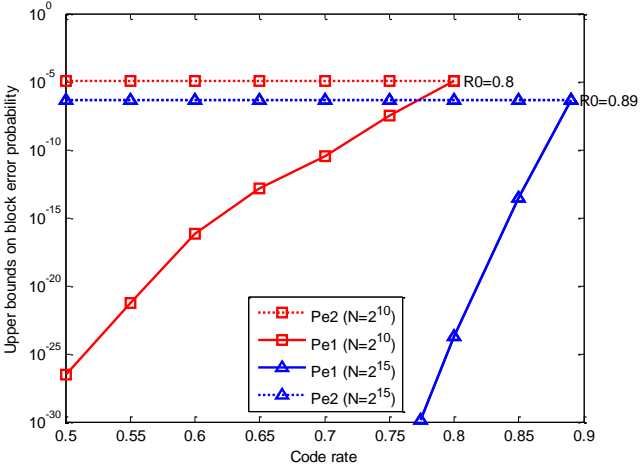

Fig. 4. Variations of $P_{e_1}$ and $P_{e_2}$ vs. rates $R \in [0.5, R_0]$ for the polar codes of lengths $N = 2^{10}, 2^{15}$ over BEC(0.05).

Moreover, as mentioned before, the intentional error vectors $\boldsymbol{e} = \boldsymbol{u}_{A^c(s)} G_{A^c(s)}$ do not affect the error correction capability of the polar codes as the polar code performance is not sensitive to the way the frozen vector $\boldsymbol{u}_{A^c(s)}$ is selected. Therefore, the error correction capability of the polar codes is fully preserved in this scheme.

### B. Key Length

In this scheme, the memory requirement of the secret key set $\mathcal{K} = \{A^c(s), IV_F, IV_S, IV_P\}$ is computed as below:

1) As mentioned before, the secret frozen set $A^c(s)$ can be saved instead of the generator matrix $G_{A(s)}$. On the other hand, the largest possible bit-channel index, i.e. $N = 1024$, might be one of the indices in $A^c(s)$. Such bit-channel index requires 11 bits to save in binary form. Hence, the upper bound on the required memory to store $A^c(s)$ is computed as $\mathcal{M}_{A^c(s)} \leq 11(N - K) = 2112$ bits.

2) The required memory to store the initial value $IV_F$ is computed as $\mathcal{M}_{IV_F} = N - K = 192$ bits.

In this scheme, a $(1024, 832)$ polar code is considered to obtain reasonable reliability and security simultaneously. On the other hand, if the nonsingular matrix $S_{832 \times 832}$ and the permutation matrix $P_{1024 \times 1024}$ are saved directly, the key length grows too large. Consequently, we attempt to apply the efficient method introduced in [22] to reduce the key length. This method is based on pseudorandom number generators, i.e. LFSRs, to reduce the memory requirements of these matrices. In this case, the short initial values $IV_S$ and $IV_P$ are saved instead of the matrices $S$ and $P$, respectively. This method takes advantage of a special type of matrices, called double-one (DBO) matrices [31], in which every single row or column contains exactly two 1s. The DBO matrix is called a DBO-1 matrix if all 1s in the matrix can be connected in a unique cycle alternately in the column and row directions. It has to be noted that all DBO-1 matrices are singular, and the rank of any $K \times K$ DBO-1 matrix is $K - 1$ according to [31].

By adding one '1' to any entry of a $K \times K$ DBO-1 matrix, we obtain the nonsingular matrix of rank $K$. Based on this interesting property, the first algorithm is introduced in [22] to construct a nonsingular matrix $S_{K \times K}$ from a relatively short seed. The input of this algorithm is an initial value, i.e. $IV_S$, of a $(2K - 2)$-bit LFSR which is applied to generate a pseudorandom sequence $r_1, r_2, \cdots, r_{2K-2}$ with 0s in the last two bits. These random bits are then used to specify the location of 1s in the $K \times K$ DBO-1 matrix. At the final stage of this algorithm, one '1' is added to any entry of the constructed $K \times K$ DBO-1 matrix. Given the property of DBO-1 matrices, the output matrix is indeed a nonsingular matrix $S_{K \times K}$. In fact, this algorithm has a one-to-one mapping from the initial value $IV_S$ to the nonsingular matrix $S_{K \times K}$.

In the second algorithm introduced in [22], a binary permutation matrix $P_{N \times N}$ is generated from an $N \times N$ DBO-1 matrix by inverting the even positions of 1s in its cycle, counting from any position. The input of this algorithm is an initial value, i.e. $IV_P$, of an $(N - 1)$-bit LFSR which is used to generate a pseudorandom sequence $r_1, r_2, \cdots, r_{N-1}$ with one '0' in the last bit. These random bits are then employed to specify the location of 1s in the permutation matrix $P_{N \times N}$. In fact, there exists a one to one mapping from the initial value $IV_P$ to the permutation matrix $P_{N \times N}$. An in-depth account of the functionality of these algorithms is beyond the scope of this paper, yet interested readers are referred to [22, 31] for a detailed description.

3) Thus, with the help of the above mentioned method, the memory requirements for storing the nonsingular matrix $S_{K \times K}$ and the permutation matrix $P_{N \times N}$ are reduced to $\mathcal{M}_{IV_S} \leq 2K - 4 = 1660$ and $\mathcal{M}_{IV_P} \leq N - 2 = 1022$ bits, respectively.

Therefore, the upper bound on the key length can be calculated as:

$$\mathcal{M}_{\mathcal{K}} = \mathcal{M}_{A^c(s)} + \mathcal{M}_{IV_F} + \mathcal{M}_{IV_S} + \mathcal{M}_{IV_P} \leq 5\text{kbit}.$$

Table II provides a comparison between the key length of the proposed cryptosystem and those of the previous RN-like cryptosystems.

TABLE II.
COMPARISON OF THE KEY LENGTHS.

| Scheme | Code | Key Length |
|---|---|---|
| Rao [10] | $C(1024,524)$ | 2 Mbits |
| Rao-Nam [6] | $C(72,64)$ | 18 kbits |
| Struik-Tilburg [11] | $C(72,64)$ | 18 kbits |
| Sun-Shieh [14] | $C(49,36)$ | 42 kbits |
| Proposed Scheme | $C(1024,832)$ | $\leq 5$ kbits |

It can be seen from the table that, although the length and dimesion of the polar code used in our scheme is much larger, the key length of our scheme is shorter than that of the previous RN-like cryptosystem.

### C. Computational Complexity

The computational complexity of the proposed scheme consists of two parts: Encryption complexity ($C_{Enc}$) and Decryption complexity ($C_{Dec}$). The encryption complexity can be expressed as:

$$C_{Enc} = C_{mul}(mS) + C_{enc}(u_{A(s)}) + C_{mul}(xP),$$

Where $C_{mul}(mS) = \mathcal{O}(K^2)$ is the number of binary operations necessary to multiply the $K$-bit message $m$ by the nonsingular matrix $S_{K \times K}$. $C_{enc}(u_{A(s)}) = \mathcal{O}(N\log N)$ is the complexity of polar encoding and $C_{mul}(xP) = \mathcal{O}(N)$ is the number of binary operations required for multiplying the $N$-bit codeword $x$ by the permutation matrix $P_{N \times N}$. In a similar vein, the decryption complexity of this scheme is defined as follows:

$$C_{Dec} = C_{mul}(yP^T) + C_{SC}(y') + C_{mul}(u_{A(s)}S^{-1}),$$

Where $C_{mul}(yP^T) = \mathcal{O}(N)$ is the number of required binary operations to perform the product of $N$-bit received vector $y$ by the transposed form of the permutation matrix $P$. Moreover, the complexity of SC decoding is $C_{SC}(y') = \mathcal{O}(N\log N)$ [2], and the number of required binary operations for multiplying the $K$-bit vector $u_{A(s)}$ by the inverse matrix $S^{-1}$ is obtained as $C_{mul}(u_{A(s)}S^{-1}) = \mathcal{O}(K^2)$.

## VI. SECURITY

Some cryptanalytic attacks such as Brute Force, Rao-Nam, Struik-Tilburg and Majority Voting have already been suggested to threat the secret key cryptosystems based on channel coding. In this section, the cryptanalytic strength of the proposed scheme against these attacks is being examined.

### A. Brute Force Attack

In the Brute Force attack, all possible keys are checked systematically until the correct key is found. However, this attack can be avoided simply if the key space is large enough. In the proposed cryptosystem, the number of parameters of the secret key set $\mathcal{K} = \{A(s), IV_F, IV_S, IV_P\}$ is obtained as explained below:

1) Since the sender selects the $K$ bit-channels randomly from all $NR_0$ good bit-channels, the number of equivalent polar codes is defined as:

$$\mathcal{N}_C(N, K) = \binom{NR_0}{K}.$$

On the other hand, the total number of $(1024, 832)$ equivalent polar codes over BEC(0.01) with $R_0 = 0.84$ is equal to $\mathcal{N}_C(1024, 832) \approx 2^{174}$. Therefore, the existing equivalent polar codes are large enough to resist against the brute force attack.

2) The number of binary nonsingular scrambling matrices $S_{K \times K}$ is equal to the number of pseudorandom sequences $r_1, r_2, \cdots, r_{2K-4}, 0, 0$ which are used to specify the locations of 1s in $K \times K$ DBO-1 matrices. Hence, the number of these binary matrices is equal to $\mathcal{N}_S = \mathcal{N}_{IV_S} \leq 2^{2K-4} - 1$. For the $(1024, 832)$ polar code, the preliminary attempts made by the adversary to find the nonsingular matrix would be impractical.

3) The number of binary permutation matrices $P_{N \times N}$ is equal to the total number of pseudorandom sequences $r_1, r_2, \cdots, r_{N-2}, 0$ used to specify the locations of 1s in $N \times N$ DBO-1 matrices. Thus, the number of these matrices is equal to $\mathcal{N}_P = \mathcal{N}_{IV_P} \leq 2^{N-2} - 1$. As a result, finding the permutation matrix is infeasible in polynomial time.

4) The number of possible $N$-bit intentional error vectors $e = u_{A^c(s)} G_{A^c(s)}$ is equal to the number of $(N - K)$-bit frozen vectors, i.e. $\mathcal{N}_e = \mathcal{N}_{IV_F} = 2^{N-K} = 2^{192}$. Hence, finding the intentional error vector by an exhaustive search is impossible.

Therefore, because of the large number of involved parameters, the exhaustive search for finding the parameters of the secret key set is likely to end in failure.

### B. Rao-Nam Attack

The Rao-Nam (RN) attack is a chosen plaintext attack operating in the following steps [6]:

1) Computing the encryption matrix $G'$ from a large set of plaintext-ciphertext $(m, c)$ pairs.
2) Recovering the message $m$ from the ciphertext $c$ using $G'$ obtained in Step 1.

The encryption algorithm of the proposed cryptosystem (Relation (4)) can be rewritten as:

$$c = mSG_{A(s)}P + u_{A^c(s)}G_{A^c(s)}P$$
$$= mG' + eP$$
$$= mG' + e', \quad (5)$$

where $G' = [g'_{ij}], i = 1, \cdots, K, j = 1, 2, \cdots, N$ is an encryption matrix and $e' = (e'_1, e'_2, \cdots, e'_N)$ is the permuted intentional error vector. Let $m_1$ and $m_2$ be two $K$-bit plaintext vectors differing only in the $i$-th, $i = 1, 2, \cdots, K$ position. Let $c_1 = m_1 G' + e'_1$ and $c_2 = m_2 G' + e'_2$ be two distinct $N$-bit ciphertext vectors achieved from the plaintexts $m_1$ and $m_2$, respectively. The difference vector of $c_1 - c_2$ is thus computed as:

$$c_1 - c_2 = (m_1 - m_2)G' + (e'_1 - e'_2) = g'_i + (e'_1 - e'_2).$$

Besides, the $i$-th row of the encryption matrix $g'_i$ is achievable through the following equation:

$$g'_i = c_1 - c_2 - (e'_1 - e'_2). \quad (6)$$

It is obvious that the Hamming weight of $(e'_1 - e'_2)$ is at most $2w_H(e')$, where $w_H(e')$ is the Hamming weight of the permuted error vector $e'$. Since the matrix $P$ is a permutation matrix, $w_H(e') = w_H(e)$. If $2w_H(e)/N \ll 1$, the difference vector $c_1 - c_2$ represents one estimate of $g'_i$. This procedure should be followed for all $i = 1, 2, \cdots, K$ to obtain the encryption matrix $G'$.

In the following, the required number of binary operations (work factor) for constructing the encryption matrix $G'$ is being computed. Let $c_j = mG' + e_jP$ and $c_k = mG' + e_kP$ be two distinct $N$-bit ciphertexts of the proposed scheme obtained from the same message $m$. The difference between $c_j$ and $c_k$ is calculated as $c_j - c_k = e'_j - e'_k$. This process should be tested until one of the values obtained for $e'_j - e'_k$ satisfies (6). Note that the complete construction of encryption matrix $G'$ must be verified, as the correctness of each vector $g'_i$ cannot be verified independently. Since the number of distinct error vectors of this scheme is equal to $\mathcal{N}_e = 2^{N-K}$, the number of all possible pairs $(e_j, e_k)$ is equal to $\binom{\mathcal{N}_e}{2} = (\mathcal{N}_e^2 - \mathcal{N}_e)/2$. In addition, the vector $g'_i$ should be computed for each of the $K$ rows of $G'$, so that the work factor of this attack is computed as $WF \geq \frac{1}{2}\left(\frac{\mathcal{N}_e^2}{2}\right)^K$. For $\mathcal{N}_e = 2^{N-K}$, the work factor is obtained as $WF = \Omega(2^{(N-K)K})$ [6]. Obviously, this attack is infeasible for the proposed cryptosystem given the fact that the number of error vectors, $\mathcal{N}_e = 2^{192}$, is too large.

Furthermore, Rao and Nam claimed that this attack can also be resisted by applying the set of intentional error vectors with a Hamming weight of $w_H(e) \approx N/2$ [6]. Later, Meijers and Tilburg [28] showed that the RN cryptosystem is vulnerable to Extended Majority Voting (EMV) attack due to the constraint on the Hamming weight of the intentional error vectors. In fact, the predefined set of error vectors has to be chosen at random. In the proposed scheme, there is not any constraint on the Hamming weight of the intentional error vectors which in turn improves the security.

*C. Struik-Tilburg Attack*

Let $\mathcal{E} = \{e_j, 1 \leq j \leq \mathcal{N}_e\}$ and $\mathcal{E}^P = \{e_jP, 1 \leq j \leq \mathcal{N}_e\}$ denote a set of distinct error vectors and their permuted error vectors, respectively. Also, consider $\mathcal{E}_\Delta = \{e_{i,j} = e_i - e_j, 1 \leq i, j \leq \mathcal{N}_e\}$ as a set of difference intentional error vectors. Similarly, $\mathcal{E}_\Delta^P = \{e_{i,j}P, 1 \leq i, j \leq \mathcal{N}_e\}$ is the set of difference permuted intentional error vectors. Since there are $\mathcal{N}_e$ distinct permuted error vectors, the set of $\mathcal{N}_e$ distinct ciphertexts is obtained as $\mathcal{C} = \{c_j = mG' + e_jP, 1 \leq j \leq \mathcal{N}_e\}$. The performance of the Struik-Tilburg (ST) attack is described in the following steps [11]:

1) First, an arbitrary message $m$ is enciphered so that a set $\mathcal{C}$ is yielded.
2) A directed labeled graph $\Gamma = (\mathcal{C}, \mathcal{E}_\Delta^P)$ is constructed whose vertices consist of $\mathcal{N}_e$ different ciphertexts and each edge from vertex $c_i$ to vertex $c_j$ is labeled as the difference permuted intentional error vector $c_i - c_j = e_{i,j}P$. Afterwards, an automorphism group $Aut(\Gamma)$ is constructed, consisting of all the permutations on $\mathcal{C}$ in which all the edges $e_{i,j}P$ remained unchanged. Hence, the cardinality of the automorphism group is $|Aut(\Gamma)| = \mathcal{N}_e$.
3) For $1 \leq i \leq K$, a message $m_i = m + u_i$ is selected where $u_i$ is a unit vector with one '1' in its $i$-th position and the rest 0s. Next, steps 1 and 2 are repeated for $m = m_i$ to construct a set of its corresponding ciphertexts $\mathcal{C}^{(i)} = \{c_j^{(i)} = m_iG' + \hat{e}_j^{(i)}P, 1 \leq i \leq K, 1 \leq j \leq \mathcal{N}_e\}$ and its directed label graph $\Gamma_i = (\mathcal{C}^{(i)}, \mathcal{E}_\Delta^P)$.
4) For $1 \leq i \leq K$, an automorphism $\Phi$ is selected randomly from the automorphism group $Aut(\Gamma)$. Then, $\Gamma_i$ is mapped on $\Gamma$ according to the selected automorphism $\Phi$. Now, $c_1^{(i)} - c_1 = m_iG' + \hat{e}_1^{(i)}P - m G' - e_1P = \hat{g}_i + \tilde{e}_1^{(i)}P$ is calculated. As there exists an automorphism $\Phi$ for which $\tilde{e}_1^{(i)} = 0$, the $i$-th row of the encryption matrix, $\hat{g}_i$, is estimated with the probability $|Aut(\Gamma)|^{-1} = \mathcal{N}_e^{-1}$.
5) Finally, using the estimated $\hat{g}_i, 1 \leq i \leq K$, the encryption matrix $\widehat{G'} = (\hat{g}_1^T, \hat{g}_2^T, \cdots, \hat{g}_K^T)$ is generated. If the solution is not correct, the steps 4 and 5 should be repeated.

As mentioned earlier, the $i$-th row of the encryption matrix, $g'_i$, can be successfully estimated with the probability $|Aut(\Gamma)|^{-1}$. In this case, the attacker should construct $|Aut(\Gamma)|^K = \mathcal{N}_e^K$ encryption matrices to finally obtain the intended encryption matrix $G'$. Therefore, obtaining the encryption matrix $G'$ requires the work factor of $O(KN\mathcal{N}_e^K)$ operations. Apparently, if the value of $|Aut(\Gamma)| = 2^{N-K}$ is large enough, this attack ends in failure. In our scheme, for a (1024,832) non-systematic polar code over $GF(2)$, there are $|Aut(\Gamma)| = 2^{192}$ intentional error vectors, implying that the ST attack is doomed to fail.

*D. Majority Voting Attack*

The Majority Voting (MV) is another kind of attack against which the cryptanalytic strength of the secret key cryptosystem based on channel coding has been analyzed [29]. An equivalent secret key cryptosystem to RN scheme is introduced in [28, 29] to be able to examine the strength of the RN scheme against MV attack.

Let $G'' = [g''_{ij}], i = 1, 2, \cdots K, j = 1, 2, \cdots, N$ be a binary $K \times N$ equivalent encryption matrix with a right inverse $(G'')^{-1}$. Let $H$ be a corresponding binary $(N - K) \times N$ parity check matrix such that $G''H^T = 0$. Moreover, $\mathcal{E} = \{e_i, 1 \leq i \leq \mathcal{N}_e\}$ is a set of $N$-bit intentional error vectors satisfying the weight property and the syndrome property of the RN cryptosystem. Finally, the syndrome error table is constructed as $\mathcal{T} = \{e_i H^T | e_i \in \mathcal{E}\}$.

*Encryption*

A $K$-bit message $m$ is encrypted into an $N$-bit ciphertext $c$ by calculating $c = mG'' + e; \; e \in \mathcal{E}$.

*Decryption*

A ciphertext $c$ is decrypted as following the steps below:
1) Compute the syndrome $s = cH^T = mG''H^T + eH^T = eH^T$. Find the corresponding error vector $e$ from the syndrome error table $\mathcal{T}$.
2) Retrieve the message $m = (c + e)(G'')^{-1}$.

The aim of the MV attack is to recover the equivalent encryption matrix $G''$ by following a number of procedures as described below [29].
1) Choose an arbitrary plaintext $m$, and compute a set of $l$ distinct encryptions of $m$, i.e. $\mathcal{C}_l = \{c_i = mG'' + e_i, 1 \leq i \leq l\}$. Let $\mathcal{E}_l = \{e_i, 1 \leq i \leq l\}$ denote the set of $l$ distinct $N$-bit intentional error vectors. Then, compute $M(\mathcal{C}_l) = M(mG'') + M(\mathcal{E}_l)$ where $M(\mathcal{C}_l)$ is an $l \times N$ matrix consisting of the ciphertexts $c_i, 1 \leq i \leq l$ in its $i$-th row, respectively. Furthermore, $M(mG'')$ is an $l \times N$ matrix where the $N$-bit vector $mG''$ is repeated in each row. Similarly, $M(\mathcal{E}_l)$ is an $l \times N$ matrix consisting of the intentional error vectors $e_i, 1 \leq i \leq l$ in its $i$-th row, respectively. The majority of the voting on each column of $M(\mathcal{C}_l)$ yields an estimate $\widehat{mG''}$, i.e. when 1s out number 0s in a column, the corresponding bit is set to '1', and otherwise to '0'.
2) Repeat the first step for a set of $K$ linearly independent messages $m_1, m_2, \cdots, m_K$ and compute $\widehat{m_1 G''}, \widehat{m_2 G''}, \cdots, \widehat{m_K G''}$.
3) Finally, obtain an estimate of the encryption matrix as $\widehat{G''} = M^{-1}(m) M(\widehat{mG''})$ where $M(m)$ is a $K \times K$ matrix consisting of the $K$-bit message $m_i, 1 \leq i \leq K$ in its $i$-th row. Besides, $M(\widehat{mG''})$ is a $K \times N$ matrix consisting of the $K$ estimates $\widehat{m_i G''}, 1 \leq i \leq K$ in its $i$-th row. This way, the estimate of the encryption matrix $G''$ is obtained and used to break the equivalent cryptosystem.

This attack requires $K$ times $l$ majority votes over $N$ coordinates. Therefore, the work factor requires an average number of $\mathcal{O}(KNl)$ bit operations [29]. Considering the worst case, i.e. $l = \mathcal{N}_e$, this attack will have a work factor of $\mathcal{O}(KN\mathcal{N}_e)$ bit operations. In the proposed cryptosystem, because of the large number of intentional error vectors, the work factor of this attack is $\mathcal{O}(2^{211})$, which is regarded as an evidence for the impracticality of the attack.

## VII. CONCLUSIONS AND FUTURE WORK

The current paper was an attempt to address the issue of applying non-systematic polar codes in the structure of secret key cryptosystems. The proposed scheme enjoys a number of advantages such as a higher security level and a shorter key length in comparison with the previous secret key cryptosystems based on channel coding. In addition, through combining the encryption and channel coding in a single step, this scheme has a potential to be implemented with a reasonable complexity which is suitable for secure high speed communications.

In this study, we employ the non-systematic polar codes due to the following reasons: (1) Existing a large family of equivalent polar codes which leads to an increase of the security level against exhaustive search attacks. (2) The special structure of the generator matrix of polar codes, because of which the scheme achieves a smaller key size. (3) The non-systematic property of polar codes, by which a specific form of intentional error vectors is obtained that can provide a higher security level against chosen plaintext attacks and a smaller key length. (4) The low complexity encoding/ decoding of the polar codes. Moreover, the construction method of these codes is simple over BECs.

The results of the investigations indicate that the security and reliability of our scheme depend on a variety of factors including the code length, code rate and secret information set. Therefore, in order to design a secure and reliable secret key scheme based on polar codes, these parameters should be selected in such a way that a suitable tradeoff is established between security and reliability.

Our future work is to apply the polar codes in the structure of McEliece public key cryptosystem. However, it has to be noted that reducing the key length of McEliece cryptosystem based on polar codes is an interesting problem.

ACKNOWLEDGEMENT

The authors would like to thank Masoumeh Kootchak Shooshtari and Behnam Mafakheri for their helpful discussions and suggestions. The authors would also like to thank Mahdi Alaghband for his careful proofreading.